\documentclass[aps,prd,preprint,groupedaddress,showpacs,showkeys]{revtex4}
\begin{document}
\title{Semiclassical (Quantum Field Theory) and Quantum (String) de Sitter Regimes: New Results}

\author{A. Bouchareb$^{1,2}$, M. Ram\'on Medrano$^{3,2}$ and N.G. S\'anchez$^{2}$}
\email{Norma.Sanchez@obspm.fr,  mrm@fis.ucm.es}
 
\affiliation{
(1) Departement de Physique, Universit\'e d' Annaba, B.P.12, El-Hadjar, Annaba 2300, Algerie.\\
(2) Observatoire de Paris, LERMA, CNRS UMR 8112,  61, Avenue de l'Observatoire, 75014 Paris, France. \\
(3) Departamento de F\'{\i}sica Te\'orica, Facultad de Ciencias F\'{\i}sicas, Universidad Complutense, 
E-28040 Madrid, Spain}

\date{\today}
\begin{abstract}

We compute the quantum string entropy $S_s(m, H)$ from the microscopic string density of states $\rho_s (m,H)$ of mass $m$ in de Sitter space-time. We find for high $m$, (high $Hm \rightarrow  c/\alpha' $), a {\bf new} phase transition at the critical string temperature $ T_{s}= (1/2\pi k_B)L_{c\ell}~c^2/\alpha'$, {\bf higher} than the flat space (Hagedorn) temperature $t_{s}$. ($L_{c\ell}= c/H$,  the Hubble constant $H$ acts at the transition as producing a smaller string constant $\alpha'$ and thus, a higher tension). $ T_s$ is the precise quantum dual of the semiclassical (QFT Hawking-Gibbons) de Sitter temperature $T_{sem}=\hbar c /(2\pi k_B L_{c\ell})$. By precisely identifying the semiclassical and quantum (string) de Sitter regimes, we find a {\bf new} formula for the full de Sitter entropy $S_{sem} (H)$, as a function of the usual Bekenstein-Hawking entropy $S_{sem}^{(0)}(H)$. For $L_{c\ell}\gg \ell_{Planck}$ , ie. for low $H\ll c/\ell_{Planck}$,  $S_{sem}^{(0)}(H)$ is the leading term, {\bf but} for high $H$ near $c/\ell_{Planck}$, a {\bf new} phase transition operates and the whole entropy $S_{sem} (H)$ is drastically  {\bf different} from the Bekenstein-Hawking entropy $S_{sem}^{(0)}(H)$. 

We compute the string quantum emission cross section $\sigma_{string}$ by a black hole in de Sitter (or asymptotically de Sitter) space-time (bhdS).  For $T_{sem~bhdS}\ll T_{s}$, (early evaporation stage), it shows the QFT Hawking emission with temperature $T_{sem~bhdS}$, (semiclassical regime). For $T_{sem~bhdS}\rightarrow T_{s}$, $\sigma_{string}$~exhibits a phase transition   into a  string de Sitter state of size $L_s = \ell_s^2/L_{c\ell}$, ($\ell_s= \sqrt{\hbar \alpha'/c}$), and string de Sitter temperature $T_s$. Instead of featuring a single pole singularity in the temperature (Carlitz transition), it features a square root {\it branch point} (de Vega-Sanchez transition). {\bf New} bounds on the black hole radius $r_g$~ emerge in the  bhdS string regime: it can become $r_g = L_s/2$, or it can reach a more quantum value, $r_g = 0.365 ~\ell_s$.

\end{abstract}
\maketitle
\section{Introduction and Results\label{sec:intro}}

The understanding of  semiclassical and quantum gravity de Sitter regimes is particularly important for several reasons:
\\
(i) the physical (cosmological) existence of de Sitter (or quasi de Sitter) stages, describing inflation at an early stage of the universe (semiclassical or quantum field theory (QFT) regime), and  acceleration at the present time  (classical regime).
\\
(ii) the need of describing de Sitter (or quasi de Sitter) quantum regimes. Besides their conceptual interest, these regimes should be relevant in the stage preceeding semiclassical inflation, their asymptotic behaviour should provide consistent initial states for semiclassical inflation, and clarify, for instance, the issue of the dependence of the observable primordial cosmic microwave fluctuations on the initial states of inflation. 
\\
(iii) the lack till now of a full conformal invariant description of de Sitter background in string theory \cite{14}. This should not be considered as an handicap for de Sitter space-time, but as a motivation for going beyond the current scarce physical understanding of string theory. The flow of consistent cosmological data (cosmic microwave background, large scale structure, and supernovae observations) place de Sitter (and quasi-de Sitter) regimes as a real part of the standard (concordance) cosmological model, \cite {17}-\cite {20}. 
\\
(iv) the results of classical and quantum string dynamics in de Sitter space-time. Besides its interesting features, solving the classical and quantum string dynamics in conformal and non conformal invariant string backgrounds it shows that the {\bf physics} is the same in the two class of backgrounds: conformal and non conformal. The mathematics is simpler in conformal invariant backgrounds, the main physics, in particular the string mass spectrum, remains the same,  \cite {11}-\cite{10},\cite{14}.

\bigskip

In this paper, we describe classical, semiclassical and quantum de Sitter regimes. A clear picture for de Sitter background is emerging, going {\bf beyond} the current picture, both for its semiclassical and quantum regimes.

A central object is  $\rho_s (m,H)$, the microscopic string density of states of mass $m$ in de Sitter background. $\rho_s (m,H)$ is derived from the string density of levels $d(n)$ of level $n$ and from the string mass spectrum $m (n, H)$ in de Sitter  background. The mass formula $m (n, H)$ is obtained by solving the quantum string dynamics in de Sitter background \cite {11}-\cite{1}. The density of levels $d(n)$ is the same in flat and in curved space-times. As a result, the mass formula $m (n, H)$ and $\rho_s (m, H)$ are {\it different} from the respective flat space-time string mass spectrum and flat space mass level density. The formulae $m(n, H)$ and $\rho_s (m, H)$ depend on the two characteristic lengths in the problem: $L_{c\ell} = c/H$ , the de Sitter radius and $\ell_s =  \sqrt{\hbar \alpha'/c}$, the fundamental string length,  or equivalently on the respective mass scales: $M_{c\ell} = c^2 L_{c\ell}/G$ and $ m_s =\ell_s / \alpha'$; relevant combinations of them emerge in the mass formula $m(n,H)$ as the de Sitter string length $L_s =\hbar/c M_s$ or string de Sitter mass $M_s = L_{c\ell}/\alpha'$. The temperature $T_s = \frac{1}{2\pi k_B}~M_s~c^2 $ emerges as a true critical string temperature for strings in de Sitter background, and moreover as the intrinsic de Sitter temperature in the quantum (string) de Sitter regime, that is in the high $H$ or high curvature de Sitter regime.
\bigskip

From $\rho_s (m, H)$, we get the string entropy $S_s^{(0)}(m, H)$. We find that a phase transition takes place at $m = M_s$,~ ie~  $T = T_s$. This is a {\bf gravitational} like phase transition: the square root {\it branch point} behaviour near the transition is analogous to the thermal self-gravitating gas phase transition of point particles \cite {13}. This is also the same behaviour as for strings in flat space-time but with the spin modes $j$ included in the  high $j$ regime, ($j\rightarrow m^2\alpha' c$), ({\it extreme} transition) \cite{5}. As pointed out in \cite {4}, this string behaviour is {\it universal}, it holds in any number of dimensions, and is similar to the Jeans's instability at finite temperature but with a more complex structure. 
\\
The transition occurs at the string de Sitter temperature $ T_{s}=  \Big(L_s/\ell_s\Big) t_s $ {\it higher} than the flat space (Hagedorn) value  $t_s $. This is so, since for high masses, the true critical string temperature in de Sitter background is $T_s$, instead of $t_s$. The flat space (Hagedorn) temperature $t_s$ is the scale temperature in the low $Hm$ regime. $H$, which acts in the sense of the string tension, does appear in the transition temperature as an "effective string tension" $(\alpha^{'}_H) ^{-1} $ : a smaller  $\alpha^{'}_H = (\hbar/c)\Big(~H \alpha'/c\Big)^2$, (and thus a {\it higher tension}). 
\\
$T_{s}$ is the precise quantum dual of the semiclassical (QFT Hawking-Gibbons) de Sitter temperature $T_{sem}=\hbar c /(2\pi k_B L_{c\ell})$, the two temperatures satisfy: $T_{s} = t_{s}^2~ T_{sem}^{-1}$. 

\bigskip

When $m \rightarrow M_s$ the string becomes ``{\it classical} ''  reflecting the classical properties of the background, $m$ becomes $M_{cl}$, (with $\alpha '$ instead of $G/c^2$), thus the string becomes the `` background ''\cite{4}. Conversely, and interestingly enough, string back reaction supports this fact:  $M_s$ is the mass of de Sitter background in its string regime, a de Sitter phase with mass $M_s$ and temperature $T_s$ is sustained by strings~\cite{1}. ($L_s$, $M_s$, $T_s$) are the {\it intrinsic} size, mass and temperature {\it of} de Sitter background in its string (high curvature) regime. 
\bigskip

By precisely identifying the semiclassical and quantum (string) de Sitter regimes, we find a {\bf new} formula for the de Sitter entropy $S_{sem} (H)$, which is a function of the usual Bekenstein-Hawking entropy $S_{sem}^{(0)}(H)$.
For $M_{cl}\gg m_{Pl}$, that is $L_{c\ell} \gg \ell_{Pl}$, i.e. low Hubble constant $H\ll c/\ell_{Pl}$, (semiclassical regime), $S_{sem}^{(0)}(H)$ is the leading term of this expression, {\bf but} for $M \rightarrow m_{Pl}$, that is  $L_{c\ell}\rightarrow \ell_{Pl}$, i.e high Hubble constant $H \rightarrow c/\ell_{Pl}$ (quantum gravity regime),  $S_{sem}^{(0)}(H)$ is sub-dominant, a gravitational phase transition operates and the whole de Sitter entropy $S_{sem}(H)$ is drastically  {\bf different}  from the semiclassical (Bekenstein-Gibbons-Hawking) de Sitter entropy $S_{sem}^{(0)}(H)$. This {\bf new phase transition} takes place at the Planck mass $m_{Pl}$, or equivalently at the Planck temperature $t_{Pl}$. Its signature is a {\it branch point square root} behaviour at the transition temperature in all space-time dimensions. 

\bigskip

We also consider the string regimes of a black hole in a de Sitter (or asymptotically) de Sitter background (bhdS). We compute the quantum string emission cross section $\sigma_{string}$ by a  Schwarschild black hole in de Sitter background. For $T_{sem~bhdS}\ll T_{s}$, (early evaporation stage), $\sigma_{string}$ shows the Hawking emission with temperature $T_{sem~bhdS}$, (semiclassical regime). For $T_{sem~bhdS}\rightarrow T_{s}$, $\sigma_{string}$ exhibits a phase transition at $T_{sem~bhdS} = T_{s}$: the  massive emission condensates into a  string de Sitter state of string de Sitter size $L_s$ and string de Sitter temperature $T_s$. Again, this is not like the flat (or asymptotically flat) space string phase transition (of Carlitz type \cite{12}, \cite{2}), but this is a de Sitter string transition: instead of featuring a single pole singularity at $T_s$, $\sigma_{string}$ features a square root branch point at $T_s$ in any D dimensions, similar to the one exhibited by  $\rho_s(m, H)$ and by the partition function of the string gaz in de Sitter space-time . This is the same behavior exhibited by the thermal selfgravitating gas of point particles (de Vega-Sanchez transition, \cite {13}) 

\bigskip

{\bf New} string bounds on the black hole emerge in the  bhdS string  regime, ie. when $T_{sem~ bhdS} = T_s$. The bhdS space-time allows an intermediate string regime, not present in the Schwarzschild black hole alone (H=0): in the asymtotically flat black hole, the black hole radius becomes $\ell_s$ in the string regime. In the asymptotically de Sitter black hole, the black hole radius $r_g$ becomes the de Sitter string size $L_s$, ($r_g = L_s/2$). If the de Sitter radius $L_{c\ell}$ reaches $L_s$, (which implies $L_s = \ell_s$ as well), then $r_g$ becomes determined by $\ell_s$, ($r_g = 0.365 ~\ell_s$).

\bigskip

This work does not make use of AdS (Anti-de Sitter space), neither of CPP's (conjectures, proposals, principles, assumed in string theory in the last years).

\section{Semiclassical de Sitter background\label{sec:dS}}

The D-dimentional de Sitter metric can be expressed in terms of the so called static coordinates as
\begin{equation}
ds^{2}=-A(r) c^2dt^{2} + A^{-1}(r) dr^2 + r^2 d\Omega_{D-2}
 \label{eq:ds}
\end{equation}
where 
\begin{equation}
A(r) = 1 - \Big(  \frac{r}{L_{c\ell}}\Big)^2 ~~,~~~~ L_{c\ell}= c H^{-1}
 \label{eq:Lcl}
\end{equation}
the horizon being located at
\begin{equation}
r =  L_{c\ell}
 \label{eq:h}
\end{equation}
$H$ is the Hubble constant and $L_{c\ell}$ the classical de Sitter length.

In the context of Quantum Field Theory (Q.F.T) in curved space time, de Sitter background has a semiclassical (Hawking-Gibbons) temperature ~\cite{7} given by
\begin{equation}
T_{sem}= \frac{\hbar}{2\pi k_B}~H = \frac{\hbar c}{2\pi k_B} ~\frac{1}{L_{c\ell}}  
\label{eq:Tsem}
\end{equation}
Eq.~(\ref{eq:Tsem}) holds in any number of space time dimensions $D$. The Hubble constant $H$ and scalar curvature $R$, or cosmological constant $\Lambda$ ($\Lambda > 0$), are related by
\begin{equation}
R = D(D-1) \frac {H^2}{c} ~~,~~ H = c ~\sqrt{\frac{2 \Lambda}{(D-1)(D-2)} }
\label{eq:H}
\end{equation}
de Sitter space time can be viewed as generated by a constant equation of state  (with positive energy density $\epsilon$ and negative pressure density $p$), satisfying
\begin{equation}
\epsilon + p =0
\label{eq:p}
\end{equation}
In $D=4$, the following relations hold
\begin{equation}
H = \frac{2}{c} ~\sqrt{\frac{2 \pi G \epsilon}{3}} ~~;~~~~\Lambda = \frac{8 \pi G}{c^4}\epsilon 
\label{eq:Hla}
\end{equation}

being $G$ the Newton gravitational constant. 

\section{ QFT Semiclassical de Sitter Entropy\label{sec:Ssem}}

In the semiclassical or Q.F.T regime, the relation between the semiclassical entropy $S_{sem}(H)$ and the density of states $\rho_{sem}(H)$ of de Sitter background is given by
\begin{equation}
\rho_{sem}(H) = e^{\frac{S_{sem}(H)}{k_B}}
\label{eq:ro}
\end{equation}

The zeroth order semiclassical de Sitter entropy is given by
\begin{equation}
S_{sem}^{(0)}(H) =  \pi k_B ~ \Bigg(  \frac{L_{c\ell}}{\ell_{pl}}\Bigg)^2 =\pi k_B ~ \Bigg(\frac{M_{cl}}{m_{pl}}\Bigg)^2
\label{eq:SH0}
\end{equation}
where
\begin{equation}
\ell_{pl} =  \sqrt{\frac{\hbar ~ G}{c^3}}~~;~~~~m_{pl}=\sqrt{\frac{\hbar ~ c}{G}}
\label{eq:lmPl}
\end{equation}
$\ell_{pl}$ and $m_{pl}$ being the Planck length and Planck mass respectively,  and $M_{cl}$ is the (classical) mass scale of de Sitter background
\begin{equation}
M_{cl} = \frac{c^2}{G}~L_{c\ell}=\frac{c^3}{G~H}~~~~~~~~(D=4)
\label{eq:Mcl}
\end{equation}
 
Eq.~(\ref{eq:SH0}) is better expressed as
\begin{equation}
S_{sem}^{(0)}(H) = \frac{1}{2} ~\frac{M_{cl}~ c^2}{T_{sem}}
\label{eq:S0}
\end{equation}
where $T_{sem}$ is the semiclassical (Gibbons-Hawking) de Sitter temperature defined  by Eq.~(\ref{eq:Tsem}). 

In terms of the classical and the semiclassical masses, $T_{sem}$ is expressed as 
\begin{equation}
T_{sem} =  \frac{c^2}{2\pi k_B} ~\frac{m_{pl}^2}{M_{c\ell}}=\frac{1}{2\pi k_B}~M_{sem}~c^2
\label{eq:TM}
\end{equation}
where
\begin{equation}
M_{sem}=\frac{m_{pl}^2}{M_{c\ell}}
\label{eq:Msem}
\end{equation}

Eq.~(\ref{eq:S0}) for the (zero order) gravitational entropy is the {\it ordinary} entropy expression for {\it any} ordinary system, where $T_{sem}$ is the Hawking temperature. The Hawking temperature $T_{sem}$ is {\it just} the Compton length of Sitter space in the units of temperature, that is, the temperature scale of the semiclassical gravity properties for which the {\it mass scale} is precisely $M_{sem}$ Eq.~(\ref{eq:Msem}). This semiclassical or intermediate energy regime interpolates between the classical and the quantum regimes of gravity. We discuss more on these regimes in Section~(\ref{sec:dual}). 
\\
As we will see in Section~(\ref{sec:dual}), the Bekenstein-Hawking entropy $S_{sem}^{(0)}(H)$ Eq.(\ref{eq:S0}), is just one term in a more general expression of the semiclassical entropy $S_{sem}(H)$ which is a function of $S_{sem}^{(0)}(H)$. For low H, that is high mass, $M_{cl} \gg m_{Pl}$, or large de Sitter radius  $L_{c\ell} >> \ell_{Pl} $, (semiclassical de Sitter regime), $S_{sem}^{(0)}(H)$ is the leading term. But for high $H$, (that is, $L_{c\ell}$  near  $\ell_{Pl} $), a gravitational phase transition operates and the whole de Sitter entropy $S_{sem}(H)$ is very different from $S_{sem}^{(0)}(H)$. The whole de Sitter entropy $S_{sem}(H)$ Eq.~(\ref{eq:ro}), as a function of the de Sitter Bekenstein-Hawking entropy $S_{sem}^{0}(H)$, will be discussed in Section~(\ref{sec:dual}). 
\section{Quantum String Entropy in de Sitter Background \label{sec:qs}}

The entropy of quantum strings in de Sitter background is defined by
\begin{equation}
\rho_{s}(m, H) = e^{\frac{S_{s}(m,H)}{k_B}}
\label{eq:rhos}
\end{equation}
 where $\rho_{s}(m, H)$  is the string density of mass levels in de Sitter space time. 

In order to derive $\rho_{s}(m, H)$, let us notice that the degeneracy $d_n(n)$ of level $n$ (counting of oscillator states) is the same in
flat and in curved space time. The differences, due to the space-time curvature, enter through the relation $m=m (n)$ of the mass spectrum. As is known, asymptotically for high $n$, the degeneracy $d_n(n)$ behaves universally as

\begin{equation}
d_n(n) = n^{-a'}~ e^{b~ \sqrt n}
\label{eq:d}
\end{equation}
where the constants $a'$ and $b$ depend on the space time dimensions and on the type of the strings; for bosonic strings:
\begin{equation}
b=2\pi\sqrt{\frac{D-2}{6}}~,~~~~~a'=\frac{D+1}{4}~~~\text{(open)}~;~~~~~~a'=\frac{D+1}{2}~~~\text{(closed)}
\label{eq:ba}
\end{equation}

For large $n$, the mass formula for quantum strings in de Sitter background is given by~\cite{8}-\cite{10}, 
\begin{equation}
\Bigg( \frac{m}{m_s} \Bigg)^2 \simeq 4 ~n  \Bigg[1- n \Bigg(\frac{m_s}{M_s} \Bigg)^2 \Bigg] ~~\text{(closed)}
\label{eq:mmsc}
\end{equation}

\begin{equation}
\Bigg( \frac{m}{m_s} \Bigg)^2 \simeq n  \Bigg[1- n \Bigg( \frac{m_s}{M_s} \Bigg)^2   \Bigg] ~~\text{(open)}
\label{eq:mmso}
\end{equation}

For $H=0$, we recover from the above expressions the flat space-time mass string spectrum. 

In  Eq.(\ref{eq:mmsc}) and  Eq.(\ref{eq:mmso}), $m_s$ is the fundamental string mass and $M_s$ the characteristic string mass in de Sitter space time ~\cite{8}-\cite{10},\cite{1},\cite{4}
\begin{equation}
m_s = \sqrt{\frac{\hbar}{\alpha'c}}\equiv~\frac{\ell_s}{\alpha'}~;~~~~M_s = \frac{L_{c\ell}}{\alpha'} = \frac{c}{H ~ \alpha'};~~~~\Bigg(\frac{m_s}{M_s} \Bigg) = \frac{\ell_{s}}{L_{c\ell}}= \frac{H}{c}\ell_{s}
\label{eq:mMs}
\end{equation}

$\alpha'$ is the fundamental string constant ($\alpha'^{-1}$ is a mass linear density), $\ell_{s}$ is the fundamental string length and $L_{c\ell}$ is given by Eq.~(\ref{eq:Lcl}). Furthermore, $M_s$ defines the quantum string de Sitter length $L_s$:
\begin{equation}
L_s = \frac{\hbar }{M_s ~c}= \frac{\ell_s^2}{L_{c\ell}}=\frac{\hbar\alpha'}{c^2}H~~,
\label{eq:Ls}
\end{equation}
and the string de Sitter temperature $T_s$:
\begin{equation}
T_s = \frac{1}{2\pi k_B} ~ M_s~c^2 = \frac{\hbar c}{2 \pi k_B} ~\frac{1}{L_s}= \frac{1}{2 \pi k_B} ~\frac{c^3}{H \alpha'} 
\label{eq:Ts}
\end{equation}

$T_s$ is the critical string temperature in de Sitter space, as it is shown in Sec.\ref{sec:partition} and Sec.\ref{sec:eBHS} below. 

The density $\rho_s(m, H)$ of mass levels and the level degeneracy $d_n(n)$ satisfy
\begin{equation}
\rho_s(m, H) ~d \Bigg(\frac{m}{m_s}\Bigg)= d_n(n) ~dn 
\label{eq:rhod}
\end{equation}
ie.
\begin{equation}
\rho_s(m, H) \simeq \frac{m}{m_s} \Bigg[ ~\frac{ d_n(n)}{g^{'}(n)} ~\Bigg]_{n=n(m)} 
\label{eq:rhof}
\end{equation}
where $\big( \frac{m}{m_s}\big)^2 \simeq g(n)$,~ $g(n)$ being read from the r.h.s of Eqs.~(\ref{eq:mmsc}) and~(\ref{eq:mmso}).

From Eqs. (\ref{eq:mmsc}) - (\ref{eq:rhof}), we derive the string mass density of levels in de Sitter background (closed strings):

\begin{equation}
\rho_s(m, H) \simeq  \frac{(m/m_s)}{ \sqrt{1-\Big(\frac{m}{M_s} \Big)^2}} 
\Bigg[ \Bigg(\frac{M_s}{m_s} \Bigg)^2 ~\frac{1}{2} \Big(1- \sqrt{1- \Big(\frac{m}{M_s}\Big)^2}~ \Big) \Bigg]^{-a'} \times
\label{eq:rhoMsc}
\end{equation}
$$\exp\Bigg\{ b \Bigg( \frac{M_s}{m_s} \Bigg) \Bigg[ \frac{1}{2} \Big(1- \sqrt{1- \Big(\frac{m}{M_s}\Big)^2} \Big) 
\Bigg]^{1/2} \Bigg\} $$

A similar formula holds for open strings. $m_s$ and $M_s$ are given by Eq.~(\ref{eq:mMs}). $\rho_s$ depends on $H$ through $M_s$ Eq.~(\ref{eq:mMs}). $M_{s}$, $(M_{s}/2)$, is the upper mass bound for closed, (open), strings.

For $H \rightarrow 0$ we recover the flat space-time solution

\begin{equation}
 \rho_s(m) \simeq  \Big( \frac{m}{m_s} \Big)^{-a} ~e^{\frac{b}{2} \big( \frac{m}{m_s} \big)}  ~~~~, ~~~~a \equiv 2a'-~1
\label{eq:rhofc}
\end{equation}
 
$\rho_s(m,H)$ Eq.~(\ref{eq:rhoMsc}) can be expressed in a more compact way as:

\begin{equation}
\rho_s (m, H)\simeq \Bigg (\frac{m}{\Delta_s M_s}\sqrt{\frac {2}{1 - \Delta_s}}\Bigg)~ 
\Bigg(\frac{M_s}{m_s}\sqrt{\frac{1 - \Delta_s}{2}}\Bigg)^{-a}~ e^{\Big(     \frac{bM_s}{m_s}\sqrt{\frac{1 - \Delta_s}{2}}\Big)} 
\label{eq:rhodm}
\end{equation}
where
\begin{equation}
\Delta_s \equiv\sqrt{1 -\Big(\frac{m}{M_s}\Big)^2}
\label{eq:deltasm}
\end{equation}

Several expressions for the exact $\rho_s(m,H)$ are useful depending on the different behaviours we would like to highligth: the  flat H=0 limit, the low mass, or the high mass behavior.\

Let us introduce the (zero order) string entropy in flat space time :
\begin{equation}
S_s^{(0)}(m) =   \frac{1}{2}~b k_B~ \Big(\frac{m}{m_s}\Big) =\frac{1}{2}~\frac{m~c^2}{t_s} 
\label{eq:Ss0c}
\end{equation}

where $t_s$ is the fundamental string temperature in flat space-time
\begin{equation}
t_s =\frac{1}{bk_B}~ m_s~ c^2 = \frac{1}{bk_B}~\frac{\hbar~c}{\ell_{s}}
\label{eq:ts}
\end{equation}

Therefore, from Eqs.~(\ref{eq:rhoMsc}) and (\ref{eq:Ss0c}), the mass density of levels $\rho_s(m, H)$ for both open and closed strings can be expressed as :

\begin{equation}
\rho_s (m, H) =\Big(  \frac{S_s^{(0)}}{k_B} \sqrt{f(x)} \Big)^{-a}~       e^{\Big(\frac{S_s^{(0)}}{k_B}\sqrt{f(x)}\Big)} ~~ 
\frac{1}{\sqrt{(1 - 4x^2)f(x)}}
\label{eq:rhoF}
\end{equation}

\begin{equation}
a=\frac{(D-1)}{2} ~~\text{(open)},~~  a=D ~~\text{(closed)}, ~~~~~~f(x) = \frac{1 - \sqrt{1 - 4 x^2}}{2x^2}
\label{eq: FX}
\end{equation}

$x$ being the dimensionless variable
\begin{equation}
x(m, H)\equiv = \frac{1}{2}\Big(\frac{m}{M_s}\Big)= 
\frac{m_s}{b M_s}\frac{S_s^{(0)}}{k_B} 
\label{eq:X}
\end{equation}

$S_s^{(0)}$ is given by Eq.(\ref{eq:Ss0c}). In terms of $\Delta_s$ Eq.(\ref{eq:deltasm}), we have:

\begin{equation}
\rho_s (m, H)\simeq \frac{1}{\Delta_s}\sqrt{\frac{1 + \Delta_s}{2}}~ 
\Bigg(\frac{S_s^{(0)}}{k_B}\sqrt{\frac{2}{1 + \Delta_s}}\Bigg)^{-a}~ e^{\Big(     \frac{S_s^{(0)}}{k_B}\sqrt{\frac{2}{1 + \Delta_s}}\Big)} 
\label{eq:rhodso}
\end{equation}
 
\begin{equation}
\Delta_s \equiv\sqrt{1-4x^2}~~ ~~,~~~~~ f(x)= \frac{2}{1+\Delta_s}
\label{eq:deltas}
\end{equation}
For small x, (small $H m\alpha '/c$), $f(x)$ can be naturally expressed as a power expansion in $x$. In particular, for $H=0$, we have $x=0$ and $f(x)=1$, and we recover the flat space time string solution: 
\begin{equation}
 \rho_s(m) \simeq  \Big( \frac{S_s^{(0)}}{k_B} \Big)^{-a} ~e^{\big( \frac{S_s^{(0)}}{k_B} \big)}
\label{eq:rhoS0f}
\end{equation}
For  $ x \ll 1 $, i.e $ m \ll M_{s} $, the corrections to the flat $(H=0)$  solution are given by: 
\begin{equation}
\rho _s(m, H))_{m\ll M_s}  \sim \left( \frac{m}
{m_s}\right) ^{-a} ~e^{\frac{b}{2}\Big(\frac{m}{m_s}\Big)\Big[~1-~\frac{1}{8}( \frac{\alpha'H m}{c})^2~ + ~O\left(\frac{m}{M_s}\right)^3~\Big]}
\label{eq:rlH}
\end{equation}

From Eqs.~(\ref{eq:rhos}), (\ref{eq:rhoF}) we can read the full string entropy in de Sitter space :

\begin{equation}
S_s(m,H) = \hat {S_s}^{(0)}(m,H) 
-a~k_B~\ln ~\big(\frac{\hat {S_s}^{(0)}(m, H)}{k_B}\big) - k_B ~\ln F(m,H)
\label{eq:SsHF}
\end{equation}
\begin{equation}
\hat {S_s}^{(0)}(m,H)\equiv S_s^{(0)}\sqrt{f(x)}~~~~~,~~~~  F\equiv \sqrt{(1 - 4x^2)f(x)}
\label{eq:SsdsHF}
\end{equation}
i.e :
\begin{equation}
S_s(m,H) = \sqrt{f(x)} ~ S_s^{(0)}  
-a~k_B~\ln \Big(\frac{\sqrt{f(x)}~S_s^{(0)}}{k_B}\Big) - k_B \ln \sqrt{f(x)}- k_B \ln~\sqrt{~1 - 4x^2~}
\label{eq:SsH}
\end{equation}

The mass domain is  $ 0 \leq  m \leq M_{s}$, ie. $ 0 \leq x \leq 1/2 $,  (which implies $   0\leq \Delta_s  \leq  1$,  ie. $1 \leq f(x)\leq 2 $). 
\\
All terms in the entropy except the first one have negative sign, (Eq.(\ref{eq:SsHF}) or Eq.(\ref{eq:SsH})).  For $\Delta_s \neq 0$, (ie. $ m \neq M_{s}$),  the entropy $S_s(m,H)$ of string states in de Sitter space is smaller than the string entropy for $H=0$. The effect of the Hubble constant is to reduce the entropy.
For low masses  $m \ll M_{s}$, the entropy is a series expansion in $(H m\alpha' /c  \ll 1)$, like a low H expansion around the flat $H=0$ solution, $S_s^{(0)}$ being its leading term. But for high masses  $m \rightarrow M_s$, that is $(H m \alpha^{'}/c) \rightarrow 1$, (i.e. $\Delta_s \rightarrow 0$), the situation is {\it very different} as we see it below.\\
Moreover, Eq.~(\ref{eq:SsHF}) for $S_s(m,H)$ allows us to write in Section~(\ref{sec:dual}) the whole expression for the semiclassical de Sitter entropy $S_{sem}(H)$, as a function of the (Bekenstein-Hawking) de Sitter entropy $S_{sem}^{(0)}(H)$.

\section{The String de Sitter Phase Transition\label{sec:sdsps}}

For  $m \sim M_s$, the string mass density of levels  Eq.~(\ref{eq:rhoMsc}) is 
\begin{equation}
\rho_s(m, H)_{m \sim M_s} \simeq \Big(\frac{M_s}{m_s}\Big)^{-a}\sqrt{\frac{M_s}{2(M_s-m)}} ~
\Bigg[ \frac{1}{2} \Big( 1 - \sqrt{\frac{2~(M_s-m)}{M_s}}  \Big)
\Bigg]^{-\frac{(1+a)}{2}}
e^{\frac{b}{\sqrt{2}} \Big(\frac{M_s}{m_s}\Big) \Big( 1 - \sqrt{\frac{M_s-m}{2~M_s}} \Big)^{\frac{1}{2}}}
\label{eq:rhoMs}
\end{equation}

Since $(M_s -  m) \ll M_s$, a power expansion of the above equation in terms of the difference $(M_s-m)/M_{s}$ yields the leading order behaviour:

\begin{equation}
\rho_s(m, H)_{m \sim M_s}  \sim \sqrt{\frac{M_s}{2~(M_s-m)}} ~~\Big(\frac{M_s}{m_s}\Big)^{-a}~
e^{\frac{b}{\sqrt 2}(\frac{M_s}{m_s})}
\label{eq:rl}
\end{equation}

Thus, for $m \sim M_s$, the entropy behaves as:
\begin{equation}
S_s(m,H)_{m \sim M_s} = k_B \ln \sqrt{\frac{M_s}{~(M_s-m)}}~-k_B\ln~2 ~+~k_B \frac{ b}{\sqrt{2}}~(\frac{M_s}{m_s})~-~ a k_B \ln~(\frac{M_s}{m_s})
\label{eq:SsMs}
\end{equation}
Or, in terms of temperature : 
\begin{equation}
S_s(T,H)_{T \sim T_s} = k_B \ln \sqrt{\frac{T_s}{~(T_s-T)}}~-k_B\ln~2 ~+~k_B \frac{ b}{\sqrt{2}}~(\frac{T_s}{t_s})~-~ a k_B \ln~(\frac{T_s}{t_s})
\label{eq:SsTs}
\end{equation}
\begin{equation}
T = \frac{1}{ 2 \pi k_B} m c^2.
\end{equation}

We see that a phase transition takes place at $m = M_s$, ie  $T = T_s$. This is a {\bf gravitational} like phase transition: the square root {\it branch point} behaviour near the transition is analogous to the thermal self-gravitating gas phase transition of point particles \cite {13}. This is also the same behaviour of the microscopic density of states and entropy of strings with the spin modes included \cite{5}. As pointed out in \cite {4}, this string behaviour is {\it universal}: this logarithmic singularity in the entropy (or pole singularity in the specific heat) holds in any number of dimensions, its origin is gravitational interaction in the presence of temperature, as Jeans's instability at finite temperature but with a more complex structure.
\\
The transition occurs at the temperature $T_{s}$ Eq.~(\ref{eq:Tsts}) {\it higher} than the (flat space) string temperature $t_{s}$:
\begin{equation}
T_s=  \frac{b}{2\pi}\Bigg(\frac{M_s}{m_s}\Bigg) t_s = \frac{b}{2\pi}\Bigg(\frac{ L_{c\ell}}{\ell_s}\Bigg) t_s 
\label{eq:Tsts}
\end{equation}
 
This is so since in de Sitter background, the flat space-time string mass $m_s$, (Hagedorn temperature $t_s$) is the scale mass, (temperature), in the {\it low} $Hm$ regime. For high masses, the critical string mass, (temperature), in de Sitter background is $T_s$, instead of $m_s$, $(t_s)$.  In de Sitter space, $H$ "pushes" the string temperature beyond the flat space (Hagedorn) value $t_s$ .
\\
\bigskip
By analogy with $t_s$, $T_s$ can be expressed as 
\begin{equation}
T_s = \frac{1} {bk_B} \sqrt{\frac{\hbar}{\alpha'_H c^2}}~~~~,~~~~ \alpha^{'}_H = \frac{\hbar}{c}\Big(\frac{2\pi}{b}~\frac{H \alpha'}{c}\Big)^2
\label{eq:TsEff}
\end{equation}
That is, $H$, which acts in the sense of the string tension, does appear in the transition temperature as an "effective string tension" $(\alpha^{'}_H) ^{-1} $ : a smaller  $\alpha^{'}_H$, (and thus a {\it higher tension}). 

The effect of $H$ in the transition is similar to the effect of angular momentum. In \cite 5  we have found a similar transition for strings in flat space-time but with the spin modes $(j)$ included: in that case, the transition occurs at a temperature $T_{j} = \sqrt{(j/\hbar)}~t_s$, {\it higher} than the Hagedorn temperature $t_{s}$, that is like an effective string constant $\alpha^{'}_j~  \equiv ~ \sqrt{\hbar/j}~ \alpha^{'}$, and thus, as a higher tension.

$\rho _s(m,H)$  Eqs.~(\ref{eq:rlH})-(\ref{eq:rl})  expresses in terms of the typical mass scales in each domain: $m_s$ (as in flat space) for low masses, $M_s$ for high masses, to which corresponds the typical number of oscillating states: $N_s \sim \mbox{Int}[ \frac{L_{cl}}{L_s}]  \sim \mbox{Int}[\frac{c^3}{\hbar \alpha ' H^2}]$, and there is the new factor $\Delta_s ^{-1}$  which is crucial for high masses. When $m > M_s$, the string does not oscillate (it inflates with the background, the proper string size is larger than the horizon  \cite{8}, \cite {15}. The meaning of the string de Sitter phase transition Eq.(\ref{eq:SsMs}) or (\ref{eq:SsTs}) is the following: when the string mass becomes $M_s$, it saturates de Sitter universe, the string size $L_s$ (Compton length for $M_s$) becomes $L_{c\ell}$, the string becomes   ``{\it classical} '' reflecting the classical properties of the background. $M_s$ is the mass of the background $M_{cl}$ Eq.~(\ref{eq:Mcl}), (with $\alpha '$ instead of $G/c^2$): for $m \rightarrow M_ s$ the string becomes  the ``{\it background} '' \cite{4}. Conversely, and interestingly enough, string back reaction supports this fact:  $M_s$ is the mass of de Sitter background in its string regime, a de Sitter phase with mass $M_s$ Eq.~(\ref{eq:mMs}) and temperature $T_s$ Eq.~(\ref{eq:Ts}) is sustained by strings \cite{1}. ($L_s$, $M_s$, $T_s$) Eqs.~(\ref{eq:mMs})-(\ref{eq:Ts}) are the
{\it intrinsic} size, mass and temperature {\it of} de Sitter background in its string (high H) regime. 


\section{Partition Function of Strings in de Sitter Background and String Bound on the Semiclassical de Sitter Temperature \label{sec:partition}}
The canonical partition function for a gaz of strings in de Sitter background is given by

\begin{equation}
\ln Z = \frac{V_{D-1}}{(2\pi)^{D-1}} \int^{M_s}_{m_0} d\Big( \frac{m}{m_s}\Big) \rho_s(m, H) ~
\int d^{D-1}k ~\ln \Bigg\{  \frac{1 + \exp \Big\{- \beta_{sem} \Big[(m^2 c^4 + \hbar^2 k^2 c^2)^{1/2}\Big] \Big\}}
{1 - \exp \Big\{- \beta_{sem} \Big[(m^2 c^4 + \hbar^2 k^2 c^2)^{1/2}\Big] \Big\}} \Bigg\} 
\label{eq:Zk}
\end{equation}

where supersymmetry has been considered for the sake of generality, $D-1$ is the number of space dimensions, $\rho_s(m, H)$ is the string density of mass levels in de Sitter space time Eq.~(\ref{eq:rhoMsc}), $\beta_{sem}= (k_B~ T_{sem})^{-1}$ with $T_{sem}$ being the semiclassical de Sitter temperature Eq.~(\ref{eq:Tsem}), ie. the Gibbons-Hawking temperature; $m_0$ is the lowest mass for which the asymptotic expression of $\rho_s(m, M_s)$ is valid, and $M_s$ Eq.~(\ref{eq:mMs}) is the upper bound for $m$ in de Sitter background. From Eq.~(\ref{eq:Zk}), we have 
\begin{equation}
\ln Z =\frac{4 V_{D-1}}{(2\pi)^{D/2}} \frac{c}{\beta_{sem}^{\frac{D-2}{2}} \hbar^{D-1}}
\sum_{n=1}^{\infty} \frac{1}{(2n-1)^{D/2}} ~
\int^{M_s}_{m_0} d\Big( \frac{m}{m_s}\Big)~ m^{D/2}~ \rho_s(m, H)~K_{D/2}
\Big( (2n-1)\beta_{sem} mc^2\Big)
\label{eq:Z}
\end{equation}

Considering the asymptotic behaviour of the Bessel function $K_{\nu}(z)\sim \Big( \frac{\pi}{2z}\Big)^2 ~e^{-z}$, and the leading order $n=1$, $(\beta_{sem} ~m~c^2 \gg 1)$, yields\\
\begin{equation}
\ln Z \simeq \frac{2 V_{D-1}}{(2\pi)^{ \frac{D-1}{2} }} ~\frac{1}{(\beta_{sem}\hbar^{2})^{\frac{D-1}{2}}}
\int^{M_s}_{m_0} d\Big( \frac{m}{m_s}\Big) \rho_s(m, H) ~m^{\frac{D-1}{2}}  ~e^{-(\beta_{sem} m c^2)}
\label{eq:Zl}
\end{equation}
 \\
Let us see the behaviors of  $\ln Z$  for low and high masses, i.e. $m \ll M_s$ and $m \sim M_s$ respectively. For $m\ll M_s$, the $\rho_s(m, H)$ leading behavior is equal to the flat space-time solution; from Eq.~(\ref{eq:Zl}),(for supersymmetric strings just multiply by a factor 2), we have for open strings:

\begin{equation}
(\ln Z)_{m\ll M_s} \sim \frac{2 V_{D-1}}{(2\pi)^{ \frac{D-1}{2}}}~
\frac{(m_s)^{\frac{D-3}{2}}}{(\beta_{sem}~\hbar^2)^{\frac{D-1}{2}}}~~
\frac{1}{(\beta_{sem}-\beta_{s})c^2}~e^{-(\beta_{sem}-\beta_{s})m_0 c^2}
\label{eq:ZmMs}
\end{equation}
\\
where $\beta_{s}=(k_B~t_s)^{-1}$,  $t_s$ being the fundamental (flat space-time) string temperature  Eq.~(\ref{eq:ts}). 
We see that for the low mass string spectrum, the canonical partition function shows a pole behaviour at $T_{sem}\rightarrow t_{s}$ , Eqs.(\ref{eq:Tsem}) and (\ref{eq:ts}). This is so, since for low  masses, (low $Hm$ regime), the string mass (temperature) scale is the flat space-time mass $m_s$, ($t_s$). This single pole (Carlitz type \cite{12}) behavior near $t_s$ is universal for any space time dimension $D$. This is the same $T_{sem}\rightarrow t_{s}$ behavior as for strings in flat space time \cite{12} and as for strings in the Schwarzchild and Kerr black holes ~\cite{2},~\cite{5}. 

For low temperatures $T_{sem} \ll t_{s}$, we recover the semiclassical (Q.F.T) non singular thermal behavior at the semiclassical (QFT) temperature $T_{sem}$:
\begin{equation}
\ln Z \simeq V_{D-1} \Big( \frac{m_s}{2 \pi \beta_{sem} \hbar^2} \Big)^{\frac{D-1}{2}}~~ e^{-\beta_{sem}m_0 c^2}
\label{eq:ZTh}
\end{equation}

From Eqs.~(\ref{eq:rhoMs}) and ~(\ref{eq:Zl}) the leading behavior of  $\ln Z$  for high masses ($m \sim M_s$) is given by 
\\
\begin{equation}
(\ln Z)_{m \sim M_s} \sim \frac{V_{D-1}}{\left(\beta_{sem}\hbar c\right)^{D-1}}~~
 \sqrt{\frac{\beta_{sem}-\beta_{sdS}} {\beta_{sem}}}
\label{eq:Zhl}
\end{equation}

\begin{equation}
(\ln Z)_{T \sim T_s}\sim V_{D-1}\left(\frac{k_B T_{sem}}{\hbar c}\right)^{D-1}~~
 \sqrt{1 - \frac{T_{sem}}{T_s}} 
\label{eq:ZhT}
\end{equation}
\\
where ~ $ \beta_{sdS} = (k_B T_s)^{-1} $,~ $T_s$ being the string de Sitter temperature Eq.~(\ref{eq:Ts}). 

Eq.(\ref{eq:Zhl}) shows a singular behavior for $\beta_{sem} \rightarrow \beta_{sdS}$ which is general for any space-time dimensions $D$; this is a square root branch point at $T_{sem}= T_s$.  That is, a phase transition takes place for $T_{sem} \rightarrow T_s$, which from Eqs. (\ref{eq:Tsem}),
(\ref{eq:TM}) and (\ref{eq:Ts}) implies $M_{c\ell}\rightarrow  m_{s}$,  $L_{c\ell}\rightarrow \ell_{s}$.

Furthermore, we see from Eq.~(\ref{eq:Zhl}) that $T_{sem}$ has to be bounded by $T_s$,  $(T_{sem} < T_s)$ . In fact, the low mass spectrum temperature condition $T_{sem}< t_{s}$ Eq.~(\ref{eq:ZmMs}), and the high mass spectrum condition  $T_{sem}< T_{s}$  Eq.~(\ref{eq:Zhl}) both imply the following upper bound for the Hubble constant $H$ (Eqs.~(\ref{eq:Lcl}),~(\ref{eq:Tsem}),~(\ref{eq:Ls}), and~(\ref{eq:Ts})):
\begin{equation}
L_{c\ell} > \ell_s, ~~\text{i.e.}, ~~H < \frac{c}{\ell_s}
\label{eq:Hb}
\end{equation}

In the string phase transition  $T_{sem} \rightarrow T_{s}$, $H$ reachs a maximum value sustained by the string tension $\alpha'^{-1}$ (and the fundamental constants $\hbar$, $c$ as well):
\begin{equation}
H_s = c ~\sqrt{\frac{c}{\alpha'\hbar}}, ~~~~(\text{i.e.},~~\Lambda_s = \frac{1}{2 {\ell_s}^2}(D-1)(D-2))
\label{eq:Hmax}
\end{equation}

The highly excited $m  \rightarrow M_s$ string gaz in de Sitter space undergoes a phase transition at high temperature $T_{sem} \rightarrow T_{s}$, into a condensate stringy state. Eqs. (\ref{eq:Hmax}) mean that the background itself becames a string state. In Section (\ref{sec:sdsps}), we showed, from the microscopic dynamical density of states $\rho_s (m, H)$, that precisely at $T = T_s$, $(m = M_s)$, the string of mass $m$ in de Sitter space undergoes a phase transition at $m = M_s$ and becomes the background itself. 
\\ 
QFT and string back reaction computations support this fact: de Sitter background is an exact solution of the semiclassical Einstein equations with the QFT back reaction of matter fields included, as well as a solution of the semiclassical Einstein equations with the string back reaction included \cite{1}: for $T_{sem} \ll T_s$, the curvature $R = R (T_{sem}, T_s)$, yields the QFT semiclassical curvature $R_{sem}$ (low H or semiclassical regime), and for $T_{sem}\rightarrow T_s$ it becomes a string state selfsustained by a string cosmological constant Eq.(\ref{eq:Hmax}). The leading term of the de Sitter curvature in the quantum regime is given by $R_s = D \: (D-1)\: c / \ell_s^2$ plus negative corrections in an expansion in powers of ($R_{sem}/R_s$) \cite{1}. The two phases: semiclassical and stringy are dual of each other in the precise sense of the classical-quantum duality \cite{1}, \cite{3},\cite{4}.
\bigskip

The results of these Sections allow also to consider the string regimes of a black hole in a de Sitter (or asymptotically) de Sitter background. This allow to study the effects of the cosmological constant on the quantum string emission by black holes, and the string bounds on the semiclassical (Gibbons-Hawking) black hole-de Sitter (bhdS) temperature~ $T_{sem~bhdS}$~, this is done in
 Sections (VII)-(IX) below. 

\section{The Semiclassical Black Hole - de Sitter Background\label{sec:BHS}}

The D-dimensional Schwarzschild - de Sitter space-time is described by the metric
\begin{equation}
ds^{2}=-a(r) ~c^2~dt^{2} + a^{-1}(r) ~dr^2 + r^2~ d\Omega_{D-2}
\label{eq:mBHS}
\end{equation}
where: 
\begin{equation}
a(r) = 1 - \frac{r_g}{r} - \Big(  \frac{r}{L_{c\ell}}\Big)^2, ~~~~
r_g =  \Bigg(  \frac{16 \pi ~G~ M}{c^2(D-2) ~A_{D-2}}\Bigg)^{\frac{1}{D-3}},~~~~
A_{D-2} =  \frac{2\pi ^{\frac{(D-1)}{2}}}{\Gamma \Big(\frac{(D-1)}{2}\Big)}
 \label{eq:am}
\end{equation}

$r_g$ being the Schwarzschild gravitational radius, and $L_{c\ell}$ is given by Eq.~(\ref{eq:Lcl}). For $D=4$, in terms of the cosmological constant $\Lambda$ Eq.~(\ref{eq:H}), one has
\begin{equation}
a(r) = 1 -  \Bigg(\frac{2 G M}{c^2}\Bigg)\frac{1}{r} - r^2 \frac{\Lambda}{3}, ~~~~~~r_g =  \frac{2 G M}{c^2},~~~~~~\Lambda = 3 \Bigg(\frac{H}{c}\Bigg)^2
\label{eq:rg4}
\end{equation}  
 
The equation $a(r)=0$ has three real solutions: $r_h$ (black hole horizon), $r_c$ (cosmological horizon), and $r_{-}~=~-(r_h+r_c)$, which has to satisfy
\begin{equation}
\frac{3}{\Lambda} = r_h^2 +r_c^2 + r_h~ r_c ~~~~\text{and}~~~~
\Bigg(\frac{2 G M}{c^2}\Bigg) \frac{3}{\Lambda} = r_h ~r_c \left( r_h +r_c\right)
\label{eq:rc}
\end{equation}

The black hole surface gravity and the cosmological surface gravity are respectively:
\begin{equation}
\mathcal{K}_{bhdS} =\frac{c^2}{2}~\Bigg \vert \frac{d a(r)}{dr} \Bigg \vert_{r=r_h} ~~, ~~~~~~
\mathcal{K}_c =\frac{c^2}{2}~\Bigg \vert \frac{d a(r)}{dr} \Bigg \vert_{r=r_c}
\label{eq:K}
\end{equation}

From the above equations we have 
\begin{equation}
\mathcal{K}_{bhdS} =\frac{c^2}{2~r_h L_{c\ell}^2} ~ (r_c - r_h)~(r_h - r_-)
\label{eq:HHS}
\end{equation}
and
\begin{equation}
\mathcal{K}_c =\frac{c^2}{2~r_c L_{c\ell}^2}~(r_c - r_h)~(r_c - r_-)
\label{eq:KcS}
\end{equation}  \\
being $r_h<r_c$, and ~$L_{c\ell}^2 = (3/\Lambda)$.~ Eq.~(\ref{eq:HHS}) can be written as well
\begin{equation}
\mathcal{K}_{bhdS} = c^2 \Bigg( \frac{r_g}{2~r_h^2}-\frac{r_h}{L_{c\ell}^2} \Bigg) 
\label{eq:Ka}
\end{equation}

From Eqs.~(\ref{eq:rc}) there is a trivial black hole horizon solution
\begin{equation}
r_h= \frac{1}{\sqrt\Lambda} = r_c~,~~~~\text{with} ~~~~~~\frac{GM}{c^2}= \frac{1}{3 \sqrt\Lambda}, 
\label{eq:rr}
\end{equation}

and a more interesting one 
\begin{equation}
r_h \simeq r_g = \frac{2GM}{c^2}~.
\label{eq:rar}
\end{equation}

From Eqs.~(\ref{eq:Ka}) and~(\ref{eq:rar}), the black hole surface gravity in the presence of $\Lambda$ is given by: 
\begin{equation}
\mathcal{K}_{bhdS} =\frac{c^2}{2~ r_g} ~ \Bigg( 1 - 2 ~\frac{r_g^2}{L_{c\ell}^2} \Bigg)~~~,
\label{eq:KH}
\end{equation}
which for $\Lambda=0$ yields the Schwarzschild surface gravity
\begin{equation}
\mathcal{K}_{bh} =\frac{c^2}{2 ~r_g}~ .
\label{eq:Krg}
\end{equation}

The Hawking black hole temperature in de Sitter space is
\begin{equation}
T_{sem~bhdS} = \frac{\hbar}{2 \pi k_B c}~~ \mathcal{K}_{bhdS} 
\label{eq:TH}
\end{equation}
which can be written as
\begin{equation}
T_{sem~bhdS} = \frac{\hbar c}{2 \pi k_B}~~\frac{1}{L_{bhdS}}
\label{eq:TBH}
\end{equation}
with
\begin{equation}
L_{bhdS} =2~ r_g \Bigg( 1 - 2 ~\frac{r_g^2}{L_{c\ell}^2} \Bigg)^{-1}
\label{eq:LH}
\end{equation}
Or, in terms of $H$ Eq.~(\ref{eq:rg4}): 
\begin{equation}
T_{sem~bhdS} = \frac{\hbar c}{2 \pi k_B}~~\frac{1}{2 r_g}~\Bigg( 1 - 2 ~ \Big(\frac{r_g~H}{c}\Big)^2\Bigg)
\label{eq:TBHS}
\end{equation}
For $H=0$, one recovers the black hole Hawking temperature
\begin{equation}
T_{sem~bh} = \frac{\hbar c}{4 \pi k_B~r_g} ~.
\label{eq:T}
\end{equation}

With these expressions for the semiclassical bhdS background we are prepared to compute the quantum emission of strings by a Schwarzschild black hole in the de Sitter background.
\section{Quantum String Emission by a Black Hole in de Sitter Background\label{sec:eBHS}}

The quantum field emission cross section $\sigma_{QFT} (k) $ of a given emitted species of particles in a mode $k$ by a black hole in de Sitter background is
given by
\begin{equation}
\sigma_{QFT}(k)=\frac{\Gamma_A}{e^{(\beta_{sem~bhdS} E(k))}-1}
\label{eq:sig}
\end{equation}

where $\Gamma_A$ is the greybody factor (absorption cross section), and for the sake of simplicity, only bosonic states have been considered ; $\beta_{sem~ bhdS}=( k_B T_{sem~bhdS} )^{-1}$, and $T_{sem~bhdS}$ is given by Eq.~ (\ref{eq:TBH}). The quantum field emission cross section of particles of mass $m$ is defined as

\begin{equation}
\sigma_{QFT}(m) = \int_{0}^{\infty} \sigma_{QFT} (k)~ d\mu (k) 
\label{eq:sm}
\end{equation}

where $d\mu (k)$ is the number of states between $k$ and $k+dk$:

\begin{equation}
d\mu(k) = \frac{2 V_{D-1}}{\Big(4\pi\Big)^{\frac{D-1}{2}}\Gamma \Big( \frac{D-1}{2} \Big) }~k^{D-2}~dk
\label{eq:mu}
\end{equation}

From Eq.~(\ref{eq:sm}) we have
$$ \sigma_{QFT}(m)=\frac{V_{D-1}~\Gamma_A}{(2 \pi)^{\frac{D-1}{2}}}
\frac{\Big( mc^2\Big)^{\frac{D-2}{2}}~}{(\beta_{sem~bhdS})^{D/2}~ (\hbar c)^{D-1}} ~~
\times $$
\begin{equation}
\sqrt{\frac{2}{\pi}}~\sum_{n=1}^{\infty} \frac{1}{n^{D/2}}~
\Big\{ n\beta_{sem~bhdS} mc^2 K_{_{D/2}} (n\beta_{sem~bhdS}mc^2) + K_{_{D/2 - 1}}  (n\beta_{sem~bhdS} mc^2) \Big\} 
\label{eq:smD}
\end{equation}

For large $m$ and the leading order $n=1$, $(\beta_{sem~bhdS}~ mc^2\gg1)$,  we obtain with the asymptotic behavior of the Bessel function $K_{\nu}$:
\begin{equation}
\sigma(m)_{QFT} \simeq \frac{V_{D-1}~\Gamma_A} {(2\pi )^\frac{D-1}{2}}~ ~
\frac{m^{\frac{D-1}{2}}} {\left(\beta_{sem~bhdS} ~\hbar^{2}\right)^ \frac{D-1}{2}}~ e^{-\beta_{sem~bhdS}~ mc^2}
\label{eq:smDl}
\end{equation}

In the string analogue model, the string quantum emission cross section, $\sigma_{string}$, is given by
\begin{equation}
\sigma_{string} \simeq  \int_{m_0}^{M_s} \rho_{s}(m, H)~\sigma_{QFT}(m)~ d\Big(\frac{m}{m_s}\Big)  
\label{eq:sD}
\end{equation}

where $\rho_{s}(m, H)$ is given by Eq.~(\ref{eq:rhoMsc}).

For $m \ll M_s$, (away from the upper mass bound and temperature $T_s$ Eq.(\ref{eq:Ts})), the $\rho_s(m, H)$ leading behaviour is given by the flat space solution $(H=0)$ Eq.~(\ref{eq:rhofc}). From Eqs.~(\ref{eq:smDl}),~(\ref{eq:sD}) and~(\ref{eq:rhofc}), (open strings), the leading contribution to the quantum string emission $\sigma_{string}$ for any D space-time dimensions is :

\begin{equation}
\sigma_{string}~(m\ll M_s) \sim \frac{V_{D-1}~\Gamma_A} {(2\pi )^\frac{D-1}{2}}~
\frac{ ~m_s^{\frac{D-3}{2}}} {\left(\beta_{sem~  bhdS}~\hbar^{2}\right)^{\frac{D-1}{2}}}~
\frac{ e^{-(\beta_{sem~bhdS}-\beta_{s})~m_0 c^2}}{\Big(\beta_{sem~bhdS} -\beta_{s}\Big)c^2}
 \label{eq:smia}
\end{equation}

For $m \ll M_s$, (which is a low $Hm$ regime), the string emission cross section shows the same singular behavior near $t_s$ as the low $Hm$ behavior of the canonical de Sitter partition function Eq.~(\ref{eq:ZmMs}), and as the quantum string emission by a (asymptotically flat) black hole ~\cite{2},~\cite{5}, here at the temperature $T_{sem~bhdS}$. This is so, since in the bhdS background, the string mass scale for low string masses (temperatures) is the Hagedorn (flat space) string temperature $t_s$. $T_{sem~bhdS} \rightarrow t_s$ is a high temperature behaviour for low $Hm \ll c/\alpha'$, $t_s$ is smaller than the string de Sitter temperature $T_s$.

For low temperatures $\beta_{sem~bhdS} \gg \beta_{s}$ we recover the semiclassical (QFT) Hawking emission at the temperature  $T_{sem~bhdS}$: 

\begin{equation}
\sigma_{string}\simeq \frac{V_{D-1}~\Gamma_A}{(2 \pi)^{\frac{D-1}{2}}}~
\frac{m_s^{\frac{D-3}{2}}}{\beta_{sem}~^{\frac{D+1}{2}}~ (\hbar c)^{D-1}}~e^{-\beta_{sem}m_0c^2}
\end{equation}

For high masses ($m \sim M_s$) we have for the $\sigma_{string}$ leading behavior : \\
\begin{equation}
\sigma_{string} ~~(m \sim M_s) \sim \frac{V_{D-1}~\Gamma_A}{\left(\beta_{sem~bhdS}~\hbar c\right)^{D-1}}~
\sqrt{\frac{\beta_{sem~bhdS}-\beta_{sdS}}{\beta_{sem~bhdS}}}
\label{eq:sim3}
\end{equation}
\begin{equation}
\sigma_{string} ~~(T \sim T_s) \sim V_{D-1}~\Gamma_A~\left(\frac{k_B T_{sem~bhdS}}{\hbar c}\right)^{D-1}
\sqrt{1~-~\frac{T_{sem~bhdS}}{T_s}}
\label{eq:siT3}
\end{equation}
\\
The black hole-de Sitter emission cross section shows a phase transition at $T_{sem~bhdS} = T_{s}$: the string emission by the black hole condensates into a de Sitter string state of string de Sitter temperature $T_s$. This is not like the flat (or asymptotically flat) space string phase transition (of Carlitz type \cite{12}, \cite{2}), but this is a de Sitter type transition. Instead of featuring a single pole singularity in $(~T~ -~T_s~)$, the  transition is a square root branch point. The branch point singular behavior at $T_{s}$ is valid for any D-dimensions and is like the one we found for the de Sitter canonical partition function Eq.~(\ref{eq:Zhl}) and for the de Sitter microscopic string density of states $\rho_s(m, H) $ Eq.~(\ref{eq:rl}) in the high $m$ (high $Hm \rightarrow c/\alpha'$) regime.  
\\ \\
The evaporation of a black hole in de Sitter space time from a semiclassical or quantum field theory regime (Hawking radiation) into a quantum string de Sitter regime (late stages), can be seen as well in the black hole decay rate. In the early evaporation stages, the  semiclassical black hole in de Sitter background decays thermally as a grey body at the Hawking temperature $T_{sem~bhdS}$ Eq.~(\ref{eq:TBHS}), with the decay rate
\begin{equation}
\Gamma_{sem} = \left| \frac{ d\ln M_{sem~bhdS} }{ d t}\right| \sim G~\left(T_{sem  ~bhdS}\right)^{3} , ~~~~~~~~M_{sem~bhdS}= 2~\pi~T_{sem~bhdS}          
\label{eq:decay}
\end{equation}

($\hbar=c=k_{B}=1$). As evaporation proceeds,$T_{sem~bhdS}$ increases until it reaches the string de Sitter temperature $T_{s}$, the black hole itself becomes an excited string de Sitter state, decaying with a string width~\cite{4},\cite{5}~ $\Gamma_{s}\sim \alpha'~T_{s}^{3}$ ,~~$(G\sim \alpha')$  into all kind of particles, with pure (non mixed) quantum radiation. The implications of the limit $T_{sem~bhdS} = T_s$ are analyzed in the Section below.

\section{String bounds for a black hole in de Sitter background\label{sec:bBHS}}

The black hole-de Sitter (bh-dS) background tends asymptotically to de Sitter space-time. Black hole evaporation will be measured by an observer which is at this asymptotic region. Asymptotically, in the Schwarzschild black hole-de Sitter space time (bh-dS), $\rho_s(m,H)$ is equal to the string mass density of states in de Sitter space time Eq.~(\ref{eq:rhoMsc}). Then, for the partition function of a gaz of strings far from the black hole in bhdS space-time, we only need to substitute  $\beta_{sem}$  in   Section (\ref{sec:partition}) by $\beta_{sem~bhdS}$,  i.e.,  substitute the de Sitter temperature $T_{sem}$ Eq.~(\ref{eq:Tsem}) by the black hole temperature in de Sitter space $T_{sem~bhdS}$ Eq.~(\ref{eq:TBHS}). With this substitution,  all results in Section (\ref{sec:partition}) hold for bhdS as well. The condition $T_{sem~bhdS} <  T_s$, (Eqs.~(\ref{eq:Lcl}),~(\ref{eq:mMs}),~(\ref{eq:LH})), yields now :
\begin{equation}
\ell_s^2 < L_{c\ell}~L_{bhdS}  
\label{eq:lLbhdS}
\end{equation}
which implies the following condition  
\begin{equation}
H~\Big[1 - 2r_g^2\left(\frac{H}{c}\right)^2\Big] < \frac{2r_{g}c}{\ell_s^2}
\label{eq:HbS}
\end{equation}

The bound is saturated ($T_{sem} = T_s$) for a  gravitational radius which satisfies 
\begin{equation}
\left(\frac{r_g}{L_{c\ell}}\right)^{2}~+~r_g \frac{L_{c\ell}}{\ell_s^2}~-~\frac{1}{2} = 0
\label{eq:rgb}
\end{equation}

yielding the physical solution

\begin{equation}
\label{eq:rgs}
r_g = \frac{1}{2}~\frac{L_{c\ell}^3}{\ell_s^2}~\Big[ -  1~+~\sqrt{ 1 + 2\left(\frac{\ell_s}{L_{c\ell}}\right)^4~}~\Big]
\end{equation}

For $L_{c\ell} \gg\ell_s $ : 
\begin{equation}
r_g \simeq \frac{1}{2}~\frac {\ell_s^2} {L_{c\ell}}~\Big[1 + O(\frac {\ell_s}{L_{c\ell}})^2 ~\Big]~~~, ie.~~~ 2 r_g \simeq \frac {H}{c}\ell_s^2 = L_s
\label{eq:rga}
\end{equation}

For $L_{c\ell}= \ell_s $ :
\begin{equation}
2r_g = 0.73~\ell_s
\label{eq:rgb}
\end{equation}

Eq.~(\ref{eq:rgs}) shows the relation between the  Schwarzschild radius and the cosmological constant Eq.~(\ref{eq:Lcl}) when $T_{sem~bhdS} = T_s$ (string regime). We see that a black hole in de Sitter space allows an intermediate string regime, not present in the Schwarzschild black hole alone $(H=0)$, since in  the bhdS background there are two characteristic string scales: $ L_s$ and $\ell_s$. In an asymtotically flat space-time, the black hole radius becomes $\ell_s$ in the string regime. In an asymptotically de Sitter space, when $T_{sem~bhdS}$ reaches $T_s$ , the black hole radius $r_g$ becomes the de Sitter string size $L_s$. If the de Sitter radius $L_{c\ell}$ reaches $L_s$, (which implies $L_{c\ell} = \ell_s$), then $r_g$ becomes determined by the scale $\ell_s$, as given by Eq.(\ref{eq:rgb}).


\section{Semiclassical (Q.F.T) and quantum (string) de Sitter regimes\label{sec:dual}}

From the microscopic string density of mass states $\rho_s(m, H)$ Secs. (\ref{sec:qs}) and (\ref{sec:sdsps}), we have shown that for  $m\rightarrow M_s$,~ i.e. $T\rightarrow T_s$, the string undergoes a phase transition into a semiclassical phase with mass $M_{cl}$ and  temperature $T_{sem}$.  Conversely, from the string canonical partition function Sec.~(\ref{sec:partition}) in de Sitter space and from the quantum string emission Sec.~(\ref{sec:eBHS}) by a black hole in de Sitter space, we have shown that for $T_{sem}\rightarrow T_s$, the semiclassical (Q.F.T) regime with Hawking-Gibbons temperature $T_{sem}$ undergoes a phase transition into a string phase at the string de Sitter temperature $T_{s}$ Eq.~(\ref{eq:Ts}). This means that in the quantum string regime, the semiclassical mass density of states $\rho_{sem}$ becomes the string mass density of states $\rho_s$ and the semiclassical entropy $S_{sem}$ becomes the string entropy $S_{s}$. Namely, a semiclassical de Sitter state, $(dS)_{sem}= (L_{c\ell}, M_{c\ell},  T_{sem}, \rho_{sem}, S_{sem})$, undergoes a phase transition  into a quantum string state $(dS)_{s}$ = $(L_{s}, m , T_{s}, \rho_{s}, S_{s})$. 

The sets $(dS)_{s}$ and $(dS)_{sem}$ are the same quantities but in different (quantum and semiclassical/classical) regimes. This is the usual classical/quantum duality but in the gravity domain, which is {\it universal}, not linked to any symmetry or isommetry nor to the number or the kind of dimensions. 
From the semiclassical and quantum de Sitter regimes $(dS)_{sem}$ and $(dS)_{s}$, we can write the full de Sitter entropy $S_{sem}(H)$, with quantum corrections included, such that it becomes the string entropy $S_s(m,H)$ Eq.~(\ref{eq:S0}) in the string regime: the full de Sitter entropy $S_{sem}(H)$ is given by

\begin{equation}
S_{sem}~(H) = \hat{S}_{sem}^{(0)}~(H)
-a~k_B~\ln ~(\frac{\hat{S}_{sem}^{(0)}~(H)}{k_B}) - k_B \ln~F(H)
\label{eq:SsemHF}
\end{equation}
where
\begin{equation}
\hat{S}_{sem}^{(0)}~(H)\equiv S_{sem}^{(0)}~(H) \sqrt{f(X)}~~~~,~~~~F(H)\equiv \sqrt{(1 - 4X^2)f(X)}    
\label{eq:Fsem}
\end{equation}
 
\begin{equation}
a=D~~ ,~~~~~f(X)= \frac{2}{1+\Delta},~~~~ \Delta \equiv\sqrt{1-4X^2}~=~ \sqrt{1 -\Big(\frac{\pi k_B}{S_{sem}^{(0)}(H)}\Big)^2} 
\label{eq:Delta}
\end{equation}
 
\begin{equation}
2 X(H)\equiv \frac{\pi k_B}{S_{sem}^{(0)}(H)}= \frac {M_{sem}}{M_{cl}}= \Big(\frac{m_{Pl}}{M_{cl}}\Big)^2   
\label{eq:X}
\end{equation}
\\
$S_{sem}^{(0)}(H)$ is the usual Bekenstein-Hawking entropy of de Sitter space Eq. (\ref{eq:S0}). $M_{cl}$ is the de Sitter mass scale Eq.(\ref{eq:Mcl}), $M_{sem}$ is the semiclassical mass Eq.(\ref{eq:Msem}). In terms of $S_{sem}^{(0)}(H)$, $S_{sem}(m,H)$ Eq.(\ref{eq:SsemHF}) reads:

\begin{equation}
S_{sem}(H) = \sqrt{f(X)} S_{sem}^{(0)}(H) 
-ak_B~\ln \Big( \sqrt{f(X)} \frac{S_{sem}^{(0)}(H)}{k_B} \Big) - k_B~\ln\sqrt{ f(X)}- k_B~\ln\sqrt{~1 - 4X^2~}
\label{eq:SsemH}
\end{equation}

$\Delta$  Eq.(\ref{eq:Delta}) with $X(H)$ Eq.(\ref{eq:X}) describes $S_{sem}(H)$ in the mass domain  $m_{pl} \leq M_{c\ell} \leq \infty$, that is,  $0\leq X \leq 1/2$. (ie. $0\leq \Delta \leq 1$). The same formula but with  $\hat{X}(H) = S_{sem}^{(0)}(H)/2\pi k_B$, instead of $X(H)$,  describes $S_{sem}(H)$ in the mass domain  $0 \leq M_{cl} \leq m_{pl}$.

\bigskip

Eq.~(\ref{eq:SsemHF}) provides the whole de Sitter entropy $S_{sem}(H)$ as a function of the Bekenstein-Hawking entropy $S_{sem}^{(0)}(H)$. $X \rightarrow 0$ means $M_{cl}\gg m_{Pl}$, that is $L_{c\ell}\gg \ell_{Pl}$ , (low $H \ll c/\ell_{Pl}$ or low curvature regime), in this case $\Delta \rightarrow 1$,  $f(X)\rightarrow 1$ and  $S_{sem}^{(0)}(H)$  is the leading term of $S_{sem}(H)$, with its logarithmic correction:
\begin{equation}
S_{sem}(H) = S_{sem}^{(0)}(H) -ak_B~\ln \Big(\frac{S_{sem}^{(0)}(H)}{k_B}\Big) 
\label{eq:SsemoH}
\end{equation}

But for {\bf high} Hubble constant, $H \sim c/\ell_{Pl}$, (ie. $M_{cl}\sim m_{Pl}$), $S_{sem}^{(0)}(H)$ is sub-dominant, a gravitational {\bf phase transition} operates and the whole entropy $S_{sem}(H)$ is drastically  {\bf different} from the Bekenstein-Hawking entropy $S_{sem}^{(0)}(H)$, as we precisely see in the Section below 

\section{The de Sitter Gravitational Phase Transition  \label{subsec:EXT}}

For $\Delta \rightarrow 0$, that is for $ M_{c\ell}\rightarrow M_{sem}$, the entropy $S_{sem}(H)$ Eq.~(\ref{eq:SsemHF}) behaves as:
\begin{equation}
S_{sem}(H)_{\Delta \sim 0} = k_B~ \ln \Delta ~+~O(1)
\label{eq:SsemMcl}
\end{equation}

The Bekenstein-Hawking entropy $S_{sem}^{(0)}(H)$ is sub-leading, O(1), in this case, ($S_{sem}^{(0)}(H)_{\Delta = 0}  = \pi k_B$). 

In terms of the mass, or temperature:
\begin{equation} 
\Delta~ =~ \sqrt{1 -  \Big(\frac{m_{Pl}}{M_{cl}} \Big)^4}~=~ \sqrt{1 - \Big(\frac{T_{sem}}{T} \Big)^2} ~~,
\label{eq:Deltasem}
\end{equation}
where
\begin{equation}
T = \frac{1}{2\pi k_B}M_{cl} c^2
\label{eq:T}
\end{equation}

and $T_{sem}$ is the semiclassical (Gibbons-Hawking) de Sitter temperature Eq.(\ref{eq:Tsem}).

In the limit $ M_{c\ell}\rightarrow M_{sem}$, which implies $ M_{cl} \rightarrow m_{Pl} $,  $S_{sem}(H)$ is dominated by

\begin{equation}
S_{sem}(H)_{\Delta \rightarrow 0} =  -~k_B~ \ln ~\Big(~\sqrt{2} \sqrt{1 - \frac{T_{sem}}{T}}~\Big)~~+~O(1)
\label{eq:SsemPl}
\end{equation}

\bigskip

This shows that a {\bf phase transition} takes place at $T \rightarrow T_{sem}$. This implies that the transition occurs for $M_{cl} \rightarrow m_{Pl}$, ie $T \rightarrow t_{Pl}$, (that is for high $H \rightarrow c/\ell_{Pl}$).  
This is a {\bf gravitational} like transition, similar to the de Sitter string transition we analysed in Section (\ref{sec:sdsps}): the signature of this transition is the square root {\it branch point} behavior at the critical mass (temperature) analogous to the thermal self-gravitating gas phase transition of point particles \cite {13}, and to the string gaz in de Sitter space.  This is also the same behaviour as that found for the entropy of the Kerr black hole in the high angular momentum $ J\rightarrow M^2G/c$ regime, (extremal transition) \cite {5}. This is {\it universal}, in any number of space-time dimensions.

\begin{acknowledgements}
A. B. acknowledges the Observatoire de Paris, LERMA, for the kind hospitality extended to him.
M. R. M. acknowledges the Spanish Ministry of Education and Science (FPA04-2602 project) for financial support, and the Observatoire de Paris, LERMA, for the kind hospitality extended to her.
\end{acknowledgements}



\end{document}